\documentclass[aps,pre,twocolumn,showkeys,nofootinbib,superscriptaddress,linenumbers]{revtex4-1}

\usepackage{utopia} 
\usepackage{amsmath}
\usepackage{amssymb}
\usepackage{graphicx}
\usepackage{verbatim}
\usepackage[colorlinks=true,citecolor=blue,linkcolor=blue]{hyperref}

\graphicspath{{figures/}}

\newcommand{\bfm}[1]{{\bf #1}}          

\newcommand{\expect}[1]{\left \langle #1 \right \rangle}                

\newcommand{\T}{\mathrm{T}}                                

\begin{document}

\title{A robust approach to estimating rates from time-correlation functions}

\author{John D. Chodera}
 \email{jchodera@berkeley.edu}
 \affiliation{California Institute of Quantitative Biosciences (QB3), University of California, Berkeley, California 94720, USA}
 
\author{Phillip J. Elms}
 \email{elms@berkeley.edu}
 \affiliation{Biophysics Graduate Group, University of California, Berkeley, California 94720, USA}
 \affiliation{Jason L.~Choy Laboratory of Single Molecule Biophysics, University of California, Berkeley, CA 94720, USA}

\author{William C. Swope}
 \email{swope@us.ibm.com}
 \affiliation{IBM Almaden Research Center, San Jose, CA 95120}

\author{Jan-Hendrik Prinz}
 \email{jan.prinz@fu-berlin.de}
 \affiliation{DFG Research Center Matheon, Freie Universit\"{a}t Berlin, Arnimallee 6, 14195 Berlin, Germany}

\author{Susan Marqusee}
 \email{marqusee@berkeley.edu}
 \affiliation{Department of Molecular \& Cell Biology, University of California, Berkeley, CA 94720, USA}
 \affiliation{California Institute of Quantitative Biosciences (QB3), University of California, Berkeley, California 94720, USA}
 \affiliation{Jason L.~Choy Laboratory of Single Molecule Biophysics, University of California, Berkeley, CA 94720, USA}

\author{Carlos Bustamante}
 \email{carlos@alice.berkeley.edu}
 \affiliation{Department of Molecular \& Cell Biology, University of California, Berkeley, CA 94720, USA}
 \affiliation{California Institute of Quantitative Biosciences (QB3), University of California, Berkeley, California 94720, USA}
 \affiliation{Department of Physics, University of California, Berkeley, CA 94720, USA}
 \affiliation{Department of Chemistry, University of California, Berkeley, CA 94720, USA}
 \affiliation{Howard Hughes Medical Institute, University of California, Berkeley, CA 94720, USA}

\author{Frank No\'{e}}
 \email{frank.noe@fu-berlin.de}
 \affiliation{DFG Research Center Matheon, Freie Universit\"{a}t Berlin, Arnimallee 6, 14195 Berlin, Germany}

\author{Vijay S. Pande}
 \thanks{Corresponding author}
 \email{pande@stanford.edu}
 \affiliation{Department of Chemistry, Stanford University, Stanford, CA 94305}

\date{\today}

\small

\begin{abstract}

While seemingly straightforward in principle, the reliable estimation of rate constants is seldom easy in practice.
Numerous issues, such as the complication of poor reaction coordinates, cause obvious approaches to yield unreliable estimates.
When a reliable order parameter is available, the reactive flux theory of Chandler allows the rate constant to be extracted from the \emph{plateau region} of an appropriate reactive flux function.
However, when applied to real data from single-molecule experiments or molecular dynamics simulations, the rate can sometimes be difficult to extract due to the numerical differentiation of a noisy empirical correlation function or difficulty in locating the plateau region at low sampling frequencies.
We present a modified version of this theory which does not require numerical derivatives, allowing rate constants to be robustly estimated from the time-correlation function directly.
We compare these approaches using single-molecule force spectroscopy measurements of an RNA hairpin.

Section: \emph{Kinetics, Spectroscopy} or \emph{Statistical Mechanics, Thermodynamics, Medium Effects}


\end{abstract}


\maketitle

\small 


The observed dynamics of complex molecular systems such as biomolecules often suggest a simple underlying behavior.
Much of chemistry and biophysics revolves around attempting to identify simple models that adequately describe the observed complex dynamics of these systems.
In many cases, stochastic conformational dynamics can be modeled to good accuracy using simple first-order phenomenological rate theory, a topic that has been extensively studied theoretically~\cite{hangi:1990:rev-mod-phys:fifty-years-after-kramers,zhou:quarterly-rev-biophys:2010:rate-theories-for-biologists}.
However, when it is necessary to estimate rates from trajectories generated by computer simulation or observed in single-molecule experiments, numerous pitfalls can frustrate the ability to extract robust, reliable, and accurate estimates of rate constants using seemingly obvious approaches.
Here, we demonstrate these pitfalls for na\"{i}ve approaches to rate estimation in single-molecule force spectroscopy for an RNA hairpin, and show how reactive flux theory~\cite{chandler:jcp:1978:reactive-flux-theory,chandler-berne:1979:jcp:reactive-flux-theory,adams-doll:surface-science:1981:multistate-reactive-flux,voter-doll:jcp:1985:multistate-reactive-flux,berne:theor-chem-acc:2000:reactive-flux-perspective} and a novel but related variation can provide robustness to sampling frequency, finite statistics, and measurement noise. 


\emph{Rate theory.}
Suppose we have a population of $N$ noninteracting molecules in solution that can occupy one of two conformational states, denoted $A$ and $B$.
Without loss of generality, we assume we are given a trajectory of some order parameter $x(t)$ that allows us to define associated occupation functions $h_A(t)$ and $h_B(t)$ for states $A$ and $B$, such that
\begin{eqnarray}
h_A(t) = \begin{cases}
1 & \mathrm{if}\:\:x(t) \le x^\ddag \\
0 & \mathrm{if}\:\:x(t) > x^\ddag \\
\end{cases} \:\: ; \:\: h_B(t) = \begin{cases}
0 & \mathrm{if}\:\:x(t) \le x^\ddag \\
1 & \mathrm{if}\:\:x(t) > x^\ddag \\
\end{cases} \nonumber
\end{eqnarray}


If there is a separation of timescales between the short relaxation time within the conformational states and the long time the system must wait, on average, in one conformational state before undergoing a transition to another state, the asymptotic relaxation behavior of an initial population of $N_A(0)$ molecules in conformation $A$ and $N_B(0)$ molecules in conformation $B$ can be described by a simple linear rate law:
\begin{eqnarray}
\frac{d}{dt} N_A(t) &=& - k_{A \rightarrow B} \, N_A(t) + k_{B \rightarrow A} \, N_B(t)  \label{equation:phenomenological-rate}
\end{eqnarray}
where $k_{A \rightarrow B}$ and $k_{B \rightarrow A}$ are microscopic rate constants.
In terms of time-dependent expectations over trajectories initiated from some initial nonequilibrium state, Eq.~\ref{equation:phenomenological-rate} is equivalent to
\begin{eqnarray}
\frac{d}{dt} \expect{h_A(t)}_{ne} &=& - k_{A \rightarrow B} \, \expect{h_A(t)}_{ne} + k_{B \rightarrow A} \, \expect{h_B(t)}_{ne} \label{equation:linear-rate-law}
\end{eqnarray}
where  $\expect{h_A(t)}_{ne}$ denotes the nonequilibrium probability of finding a given molecule in conformation $A$ at time $t$ given that the fraction of molecules that were initially in conformation $A$ was $\expect{h_A(0)}_{ne} = N_A(0) / N$.

Were Eq.~\ref{equation:linear-rate-law} to govern dynamics at all times, the expected fraction of molecules in conformation $A$ as a function of time would be given by an exponential decay function,
\begin{eqnarray}
\expect{h_A(t)}_{ne} &=& \expect{h_A} + [\expect{h_A(0)}_{ne} - \expect{h_A}] \, e^{- k t}  \label{equation:first-order-rate-solution},
\end{eqnarray}
where $\expect{h_A}$ denotes the standard equilibrium expectation of $h_A$, giving the equilibrium fraction of molecules in conformation $A$. 
The quantity $k \equiv k_{A \rightarrow B} + k_{B \rightarrow A}$ denotes the \emph{phenomenological rate constant} because it is the effective rate that dominates the observed exponential asymptotic relaxation decay behavior.
It is the estimation of this quantity, $k$, that will be our primary concern.

If the system were purely two-state, such that Eq.~\ref{equation:first-order-rate-solution} held for \emph{all} time $t > 0$, a number of na\"{i}ve approaches to estimation of the phenomenological rate constant $k$ from observed trajectory data would yield useful rate estimates.
For example, given an observed trajectory $x(t)$, we could simply compute the number of times $n_c$ the dividing surface $x^\ddag$ was crossed in either direction in total trajectory time $t_\mathrm{obs}$, estimating the $k$ rate by,
\begin{eqnarray}
k_\mathrm{crossings} &\approx& \frac{n_c}{t_\mathrm{obs}} \label{equation:rate-estimate-from-crossings} .
\end{eqnarray}
Alternatively, we could partition the trajectory into segments where the system remains on one side of $x^\ddag$ in each segment, and estimate the \emph{mean lifetime} $\tau$ of these segments, from which the rate $k$ is estimated by, 
\begin{eqnarray}
k_\mathrm{lifetime} &\approx& \tau^{-1} \label{equation:rate-estimate-from-lifetimes} .
\end{eqnarray}
Both approaches will yield rate estimates that converge to the true rate $k$ as $t_\mathrm{obs} \rightarrow \infty$ when $x$ provides a perfect reaction coordinate for a perfectly two-state system, in that $x^\ddag$ correctly divides the two conformations states that interconvert with first-order kinetics.

However, when considering trajectories obtained from computer simulations or single-molecule experiments with imperfect dividing surfaces, these na\"{i}ve approaches can lead to substantially erroneous estimates.
First, we do not expect Eq.~\ref{equation:first-order-rate-solution} to hold for short times $t < \tau_\mathrm{mol}$, where $\tau_\mathrm{mol}$ is the timescale associated with relaxation processes that damp out \emph{recrossings} that occur due to imperfect definition of the separatrix between the reactant and product states~\cite{chandler:jcp:1978:reactive-flux-theory,chandler-berne:1979:jcp:reactive-flux-theory,berne:theor-chem-acc:2000:reactive-flux-perspective}.
An alternative view of this is that the observed coordinate $x$ might function as a good order parameter, in that it allows the conformational states to be well-resolved at extreme values of $x$, but a poor reaction coordinate, in that both conformational states are populated in some region near the optimal dividing surface $x^\ddag$~\cite{chodera-pande:2011:prl:single-molecule-pfold,hummer:2004:jcp:transition-paths,best-hummer:2005:pnas:reaction-coordinates} (which is optimal in that it minimizes the rate estimate in a variational sense~\cite{truhlar-garrett:annu-rev-phys-chem:1984:variational-tst}).
The rate estimates from Eqs.~\ref{equation:rate-estimate-from-crossings} and \ref{equation:rate-estimate-from-lifetimes} will therefore overestimate the number of crossings or underestimate the state lifetimes, instead converging to the transition state theory rate estimate $k_\mathrm{TST}$ that gives the instantaneous flux across the dividing surface,
\begin{eqnarray}
k_\mathrm{TST} &\equiv& \left. \frac{d}{dt} \frac{\expect{h_A(0) h_B(t)}}{\expect{h_A}} \right|_{t=0} \label{equation:tst-estimate} ,
\end{eqnarray}
and hence overestimating the true rate $k$.
Additionally, if the observed trajectories are not continuous, but instead consist of discrete observations made with a sampling resolution $\Delta t$, additional issues develop.
As the sampling interval $\Delta t$ increases, some crossing of the dividing surface $x^\ddag$ will be missed, and the perceived lifetimes of states will be increased, having the \emph{opposite} effect of a poor reaction coordinate in \emph{diminishing} the rate estimates of Eqs.~\ref{equation:rate-estimate-from-crossings} and \ref{equation:rate-estimate-from-lifetimes}.
As a result, it can be difficult to predict whether the overall result is an underestimate or overestimate of the true rate $k$.
An example illustrating these effects for a model system where the true rate is known is given in the \emph{Supplementary Material}.

\begin{figure}[tbp]
\resizebox{\columnwidth}{!}{\includegraphics{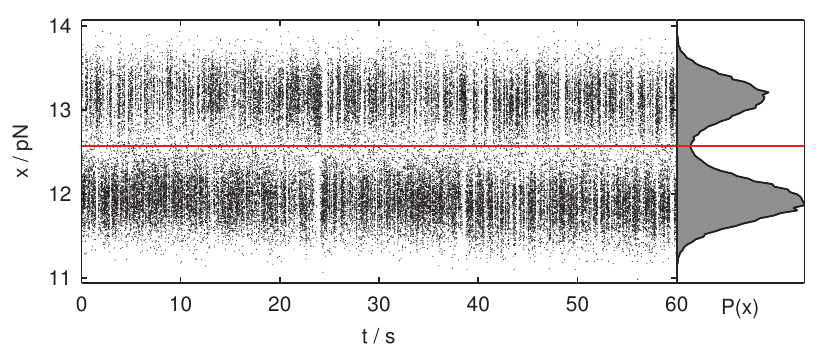}} 
\caption{\label{figure:p5ab-force-trace} {\bf Force trace of p5ab RNA hairpin in a stationary optical trap.}
A 60-second observation is shown, where the force history $x(t)$ recorded at 50 kHz and subsampled to 1 kHz is plotted.
A histogram of the observed force values is shown as $P(x)$ to the right.
The red line indicates the optimal dividing surface for rate calculations, $x^\ddag \approx 12.57$ pN.
} 
\end{figure}

\emph{Application of na\"{i}ve rate estimators to single-molecule data.}
To understand how these pathologies can affect real measurements, we examined the behavior of the p5ab RNA hairpin in an optical trap under passive conditions.
This hairpin has been the subject of previous single-molecule force spectroscopy studies~\cite{liphardt:science:2001:p5ab,wen:biophys-j:2007:p5ab,elms:2010:force-feedback-experiments}, and exhibits apparent two-state kinetics as the hairpin folds and unfolds under an external biasing force.  
The force trace $x(t)$ is shown in Fig.~\ref{figure:p5ab-force-trace}, and reports the instantaneous force on the optically trapped bead along the bead-bead axis; for a harmonic trap, this force is linearly proportional to the displacement of the bead from the center of the trap, and hence the bead-to-bead extension.
As the hairpin folds, the bead-to-bead distance contracts, increasing the applied force as the polystyrene bead conjugated to the end of the polymer moves away from the center of the optical trap. 
At the stationary trap position used for data collection, the hairpin makes many transitions between the two states resolvable from the measured force in the 60-second trajectory, populating each state nearly equally (Fig.~\ref{figure:p5ab-force-trace}). 
Data was collected at 50 kHz using a dual-beam counter-propagating optical trap~\cite{smith:methods-enzym:2003:minitweezers,bustamante-smith:2006:minitweezers-patent}, a high sampling rate far above the corner frequency for bead response under these conditions, as previously published~\cite{elms:2010:force-feedback-experiments}.
To examine the dependence on sampling interval $\Delta t$, the data was also subsampled to 1 kHz, a frequency found to be below the corner frequency of the bead, such that the bead velocity has decorrelated between sequential observations due to hydrodynamic interactions~\cite{elms:2010:force-feedback-experiments}.

\begin{figure}[tbp]
\resizebox{\columnwidth}{!}{\includegraphics{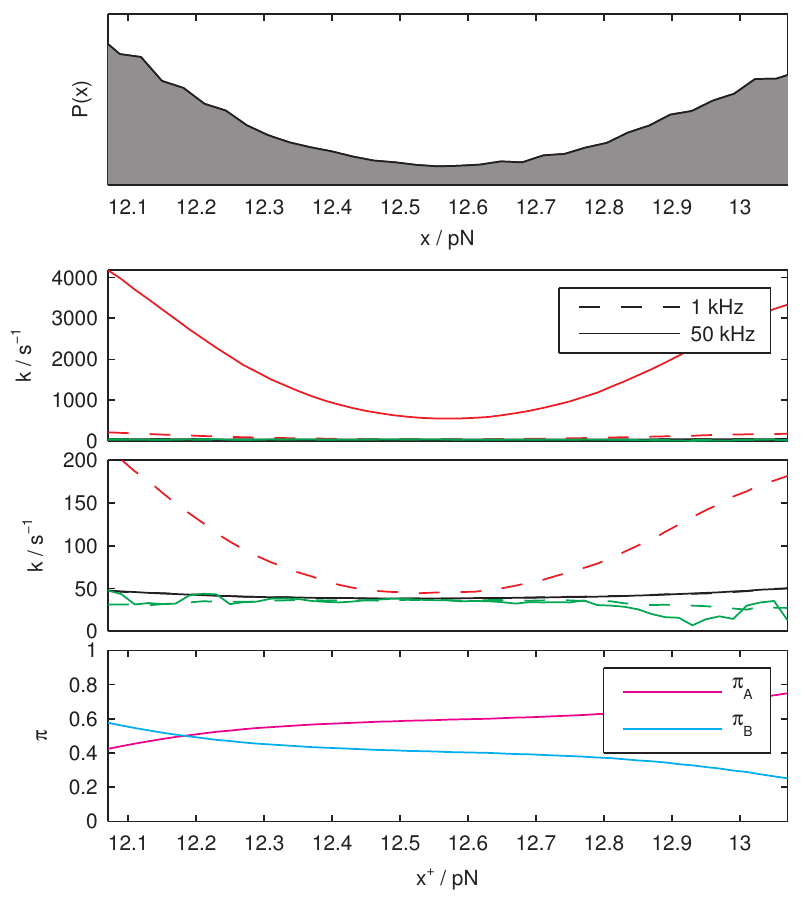}} 
\caption{
\label{figure:p5ab-dividing-surface} 
{\bf Dependence of rate estimates on dividing surface.}
\emph{Top:} Histogram of observed forces near transition region between conformational states.
\emph{Upper middle:} Rate estimate from crossing rate $k_\mathrm{crossing}$ (red lines), reactive flux rate estimated $k_\mathrm{RF}(\tau)$ near plateau time of $\tau = 3$ ms (green lines), and implied rate $k_\mathrm{im}(\tau)$ evaluated at $\tau = 60$ ms (black lines), estimated from 1 kHz data (dashed lines) or 50 kHz data (solid lines) as a function of dividing surface $x^\ddag$ choice.
\emph{Lower middle:} Same as upper middle, but close-up view of rate estimates below 200 s$^{-1}$.
\emph{Bottom:} Estimates of equilibrium probabilities $\pi_A$ and $\pi_B$ estimated from 1 kHz data as a function of dividing surface placement $x^\ddag$.  
(Estimates of $\pi_A$ and $\pi_B$ from 50 kHz data are visually indistinguishable from 1 kHz estimates.)
}
\end{figure}

The rate constant was estimated using the na\"{i}ve crossing rate $k_\mathrm{crossings}$ (Eq.~\ref{equation:rate-estimate-from-crossings}) as a function of the dividing surface choice $x^\ddag$, and plotted in Fig.~\ref{figure:p5ab-dividing-surface} (middle upper and lower panels, red lines).
Two issues are quickly discerned: 
First, near the optimal choice of dividing surface ($x^\ddag \sim 12.57$ pN), the estimated rate $k_\mathrm{crossing}$ differs greatly depending on whether the 1 kHz data (red dashed line) or 50 kHz data (red solid line) were used to compute the rate estimate, yielding disparate estimates of 45.4 s$^{-1}$ and 552 s$^{-1}$, respectively.
Second, as the dividing surface is perturbed slightly, the rate estimate for either sampling rate changes rapidly.
Both properties are highly undesirable, as practical estimators of the rate should yield results insensitive to the sampling rate and exact placement of dividing surface.

\emph{Reactive flux theory.}
To deal with the problems inherent in using an imperfect reaction coordinate or dividing surface, Chandler (and subsequent workers) demonstrated how the phenomenological rate could be recovered through the use of time-correlation functions, proposing the \emph{reactive flux} $k_\mathrm{RF}(t)$ be computed~\cite{chandler:jcp:1978:reactive-flux-theory,chandler-berne:1979:jcp:reactive-flux-theory,adams-doll:surface-science:1981:multistate-reactive-flux,voter-doll:jcp:1985:multistate-reactive-flux} to estimate $k$,
\begin{eqnarray}
k_\mathrm{RF}(t) &=& -\frac{d}{dt} \frac{\expect{\delta h_A(0) \, \delta h_A(t)}}{\expect{\delta h_A^2}} \label{equation:reactive-flux-correlation},
\end{eqnarray}
where $\delta h_A(t) \equiv h_A(t) - \expect{h_A}$ is the instantaneous deviation from the equilibrium population for some trajectory $x(t)$.
The reactive flux function $k_\mathrm{RF}(t)$ measures the flux across the boundary between $A$ and $B$ that is \emph{reactive}, in the sense that the system has crossed a dividing surface placed between $A$ and $B$ at time zero and is located on the \emph{product} side of the boundary at time $t$.
The reactive flux is bounded from above by the transition state theory rate estimate $k_\mathrm{TST}$, the instantaneous flux across the boundary, because recrossings back to the reactant state will diminish the reactive flux; $k_\mathrm{RF}(t)$ becomes identical to $k_\mathrm{TST}$ as $t \rightarrow 0^+$~\cite{chandler:jcp:1978:reactive-flux-theory}.
At $t$ larger than some $\tau_\mathrm{mol}$, thermalization processes will cause recrossings to die out, and the molecule will be captured either in its reactant or product states and remain there for a long time.
As a result, the asymptotic rate constant (whose existence requires the presupposed separation of timescales) is only obtained at $\tau_\mathrm{mol} < t \ll \tau_\mathrm{rxn}$, where $k_\mathrm{RF}(t)$ reaches a \emph{plateau value}.
$k_\mathrm{RF}(t)$ subsequently decays to zero at $t \gg  \tau_\mathrm{rxn}$ with a time constant of $\tau_\mathrm{rxn} = 1/k$~\cite{chandler:jcp:1978:reactive-flux-theory,chandler-berne:1979:jcp:reactive-flux-theory}.
Subsequent work extends these concepts to the case of multiple conformational states~\cite{adams-doll:surface-science:1981:multistate-reactive-flux,voter-doll:jcp:1985:multistate-reactive-flux}.

\begin{figure}[tbp]
\resizebox{\columnwidth}{!}{\includegraphics{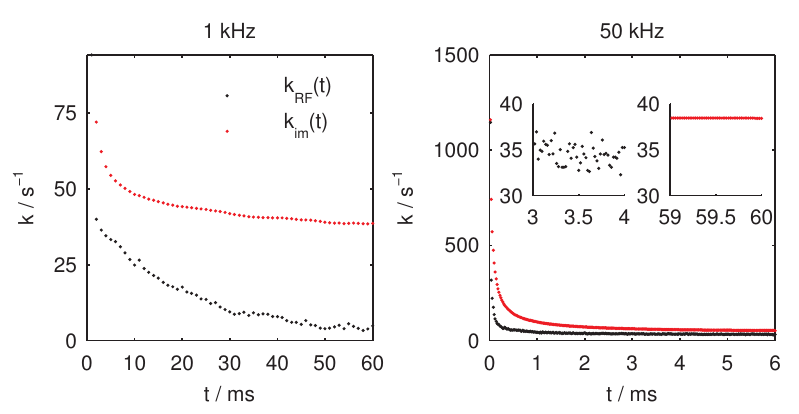}} 
\caption{\label{figure:p5ab-hairpin} {\bf Reactive flux and implied rates from p5ab hairpin single-molecule force trajectory.}
The implied rate $k_\mathrm{im}(t)$ (red) and reactive flux rate correlation function $k_\mathrm{RF}(t)$ (black) are computed for the optimal dividing surface $x^\ddag \approx 12.57$ pN for 1 kHz (\emph{left}) and 50 kHz (\emph{right}).
Close-up views compare the scatter in the rate estimates in the plateau region (3--4 ms) and long correlation times (59--60 ms) for 50 kHz data (\emph{right insets}).
} 
\end{figure}

\emph{Application of reactive flux theory to single-molecule data.}
We computed the reactive flux $k_\mathrm{RF}(t)$ from this force trajectory for both 1 kHz and 50 kHz sampling frequencies, using one-sided finite-differences to estimate the derivative in Eq.~\ref{equation:reactive-flux-correlation}. 
When estimated from 50 kHz data (Fig.~\ref{figure:p5ab-hairpin}, right), the reactive flux $k_\mathrm{RF}(t)$ smoothly stabilizes to $\sim 36$ s$^{-1}$ after a transient time of $\tau_\mathrm{mol} \approx 3$ ms.
This is the \emph{plateau time} $t = 3$ ms for which $k_\mathrm{RF}(t) \approx k$, with $t_\mathrm{mol} < t \ll \tau_\mathrm{rxn}$, where $\tau_\mathrm{rxn} \approx 28$ ms.
When the reactive flux $k_\mathrm{RF}(t)$ is evaluated at $t = 3$ ms for various choices of dividing surface $x^\ddag$ in this transition region (Fig.~\ref{figure:p5ab-dividing-surface}, middle panels), the reactive flux rate is indeed insensitive to the choice of $x^\ddag$ at both 50 kHz (solid green line) and 1 kHz (dashed green line) sampling frequencies.
In this respect, the reactive flux approach provides a much more robust way to estimating rates than the na\"{i}ve estimators of Eqs.~\ref{equation:rate-estimate-from-crossings} and \ref{equation:rate-estimate-from-lifetimes}.
To our knowledge, this represents the first time this theory has been applied to single-molecule experiments.

\emph{Implied rate theory.}
The universal application of the reactive flux approach to rate estimation from single-molecule and computer experiments still presents a number of practical difficulties.
Because the correlation function is estimated from a trajectory $x(t)$ sampled with discrete time resolution $\Delta t$, computation of the time derivative in Eq.~\ref{equation:reactive-flux-correlation} by finite-difference methods can often introduce unacceptably large amount of noise in the resulting estimate of $k_\mathrm{RF}(t)$ (Fig.~\ref{figure:p5ab-hairpin}, right inset, black dots).
Alternatively, the correlation function $\expect{\delta h_A(0) \delta h_A(t)}$ could be smoothed by fitting a polynomial to produce a continuous estimate of the derivative, but this introduces a bias that is difficult to quantify.
Additionally, if the reaction timescale $\tau_{rxn}$ is not very long compared to the observation interval $\Delta t$, then the plateau region where $k_\mathrm{RF}(t)$ is identical to the rate may be small and difficult to detect before $k_\mathrm{RF}(t)$ decays to zero.
This can be seen in the reactive flux $k_\mathrm{RF}(t)$ estimated from the 1 kHz data (Fig.~\ref{figure:p5ab-hairpin}, left, black dots), where the plateau region near 3--4 ms is relatively narrow and difficult to detect, and the $k_\mathrm{RF}(t)$ falls (decaying as $k e^{-k t}$) as $t$ reaches times comparable to $\tau_\mathrm{rxn}$.
Lastly, while alternative expressions to Eq.~\ref{equation:reactive-flux-correlation} exist where the \emph{velocity} normal to the separatrix at the time of barrier crossing is utilized instead of a time derivative of the empirical correlation function~\cite{chandler:jcp:1978:reactive-flux-theory,chandler-berne:1979:jcp:reactive-flux-theory}, it is difficult to compute this velocity for complex dividing surfaces in computer simulations, and difficult to measure experimentally in single-molecule experiments.


We propose an alternative approach, similar in spirit to reactive flux but more closely related to the rate theories used in constructing Markov state models from molecular simulations~\cite{swope-pitera-suits:jpcb:2004:markov-models-theory,hummer:njp:2005:rate-matrix-sampling,chodera:mms:2006:markov-models,noe:jcp:2011:msm-review}, that avoids the need to compute the time derivative of the correlation function in Eq.~\ref{equation:reactive-flux-correlation}.
Instead, we estimate the rate $k_\mathrm{im}(t)$ \emph{implied} by the state-to-state transition probabilities observed for a given observation interval $t$---referring to this quantity as the \emph{implied} rate constant.
As with the reactive flux, for times $t$ where $\tau_\mathrm{mol} < t \ll \tau_\mathrm{rxn}$, the phenomenolgical rate constant (if it exists, by virtue of a separation of timescales) is recovered by $k_\mathrm{im}(t)$, but our modified estimator provides a much larger plateau for times $t > \tau_\mathrm{mol}$ where a usable rate estimate can be extracted.

As before, if a separation of timescales exists, relaxation behavior for times $t >\tau_\mathrm{mol}$ is defined in terms of first order rate equations (Eq.~\ref{equation:linear-rate-law}), here recast in matrix form,
\begin{eqnarray}
\frac{d}{dt} \bfm{p}(t) &=& \bfm{K} \, \bfm{p}(t) \label{equation:rate-matrix-evolution}
\end{eqnarray}
where $\bfm{p} = [p_A(t) \: p_B(t)]^\T$,  $p_A(t) = \expect{h_A(t)}_{ne}$ and $p_B(t) = \expect{h_B(t)}_{ne}$ denote the nonequilibrium occupation probabilities of states $A$ and $B$ at time $t$, and $\bfm{K}$ is the matrix of rate constants
\begin{eqnarray}
\bfm{K} &=& \left[ \begin{array}{rr}
-k_{A \rightarrow B} & k_{B \rightarrow A} \\
k_{A \rightarrow B} & - k_{B \rightarrow A}
\end{array}\right] .
\end{eqnarray}
The eigenvalues of $\bfm{K}$ are $\lambda_1 = 0$, reflecting conservation of probability mass, and $\lambda_2 = - (k_{A \rightarrow B} + k_{B \rightarrow A}) = - k$, which governs the recovery toward equilibrium populations $\pi_A$ and $\pi_B$ at the phenomenological relaxation rate $k$.

The solution to Eq.~\ref{equation:rate-matrix-evolution} (corresponding to Eq.~\ref{equation:first-order-rate-solution}) is given by
\begin{eqnarray}
\bfm{p}(t) &=& e^{\bfm{K} t} \, \bfm{p}(0) = \bfm{T}(t) \, \bfm{p}(0)
\end{eqnarray}
where $e^{\bfm{A}} \equiv \sum_{n=0}^\infty \bfm{A}^n / n!$ is the formal matrix exponential and $\bfm{T}(t)$ can be identified as the column-stochastic \emph{transition probability matrix} whose elements $T_{ji}(t)$ give the conditional probability of observing the system in conformation $j$ at time $t$ given that it was initially in conformation $i$ at time $0$.

The elements of $\bfm{T}(t)$ for a given observation interval $t$ are conveniently given in terms of the time-correlation function,
\begin{eqnarray}
T_{ji}(t) &\equiv& \frac{\expect{h_i(0) \, h_j(t)}}{\pi_i} \label{equation:transition-matrix}
\end{eqnarray}
where the stationarity and time-reversal symmetry of physical systems at equilibrium ensures that $\expect{h_i(0) \, h_j(t)} = \expect{h_j(0) \, h_i(t)}$, and $\pi_i$ is the equilibrium probability of state $i$.

For $t > \tau_\mathrm{mol}$, we have $\bfm{T}(t) \approx e^{\bfm{K} t}$ for a constant matrix $\bfm{K}$, but this will not hold for $t < \tau_\mathrm{mol}$.
Instead, we can establish a one-to-one correspondence between $\bfm{T}(t)$ and the rate matrix $\bfm{K}_{im}(t)$ it \emph{implies} for any $t$,
\begin{eqnarray}
\bfm{T}(t) = e^{\bfm{K}_{im}(t) \, t}  \:\: \Leftrightarrow \:\: \bfm{K}_{im}(t) = t^{-1} \log \bfm{T}(t) \label{equation:definition-of-implied-rate-matrix} ,
\end{eqnarray}
where the logarithm denotes the matrix logarithm.
Assuming a phenomenological rate constant $k$ \emph{exists}, all $\bfm{K}_{im}(t) \approx \bfm{K}$ for $t > \tau_\mathrm{mol}$.

Because of their relationship through the exponential (Eq.~\ref{equation:definition-of-implied-rate-matrix}), $\bfm{T}(t)$ and $\bfm{K}_{im}(t)$ share the same eigenvectors $\bfm{u}_k$, and their respective eigenvalues $\mu_k(t)$ and $\lambda_k(t)$ are simply related~\cite{hummer:jpcb:2008:coarse-master-equations},
\begin{eqnarray}
\mu_k(t) &=& e^{\lambda_k(t) \, t} . 
\end{eqnarray}
The implied rate constant $k_\mathrm{im}(t)$ for observation time $t$ can be obtained from the nonzero eigenvalue of $\bfm{K}_{im}(t)$, 
\begin{eqnarray}
k_\mathrm{im}(t) &=& - \lambda_2(t) = - t^{-1} \ln \mu_2(t) \label{equation:implied-rate-in-terms-of-eigenvalue}
\end{eqnarray}
where $\mu_2(t) = 1 - (T_{AB}(t) + T_{BA}(t))$.
Using Eq.~\ref{equation:transition-matrix} and some algebra, we find $\mu_2(t)$ can be written,
\begin{eqnarray}
%
\mu_2(t) &=& \frac{\expect{\delta h_A(0) \delta h_A(t)}}{\expect{\delta h_A^2}} \label{equation:second-eigenvalue} ,
\end{eqnarray}
which is simply the normalized fluctuation autocorrelation function for the indicator function $h_A$ for state $A$ (or, equivalently, for state B).
$\mu_2(t)$ therefore takes the value of unity at $t = 0$ and decays to zero at large $t$.

Combining Eqs.~\ref{equation:implied-rate-in-terms-of-eigenvalue} and \ref{equation:second-eigenvalue} gives the expression for the implied rate estimate $k_\mathrm{im}(t)$ of the phenomenological rate $k$,
\begin{eqnarray}
k_\mathrm{im}(t) &=& - t^{-1} \ln \frac{\expect{\delta h_A(0) \delta h_A(t)}}{\expect{\delta h_A^2}} \label{equation:implied-rate} ,
\end{eqnarray}
which is the main result of this paper.

In the limit $t \rightarrow 0^+$, $k_\mathrm{im}(t)$ reduces to the transition state theory estimate $k_\mathrm{TST}$,
\begin{eqnarray}
\lim_{t \rightarrow 0^+} k_\mathrm{im}(t) &=& - \left. \frac{d}{dt} \frac{\expect{\delta h_A(0) \delta h_A(t)}}{\expect{\delta h_A^2}} \right|_{t = 0} = k_\mathrm{TST} ,
\end{eqnarray}
just as for the reactive flux rate (Eq.~\ref{equation:reactive-flux-correlation})~\cite{chandler:jcp:1978:reactive-flux-theory,chandler-berne:1979:jcp:reactive-flux-theory}.
Similarly, the true phenomenological rate $k$ is given by the long-time limit of $k_\mathrm{im}(t)$:
\begin{eqnarray}
k &=&  \lim_{t \rightarrow \infty} k_\mathrm{im}(t) = \lim_{t \rightarrow \infty} - t^{-1} \ln \frac{\expect{\delta h_A(0) \delta h_A(t)}}{\expect{\delta h_A^2}}
\end{eqnarray}
However, when estimating the phenomenological rate through this expression, evaluation of the correlation function should be for some $t \ll \tau_{rxn} = k^{-1}$, as the statistical error in the estimate of $k_\mathrm{im}(t)$ grows with $t$ (see Appendix).


When there is a separation of timescales such that $\tau_\mathrm{mol} \ll \tau_\mathrm{rxn}$, such that a phenomenological rate exists, we can see that $k_\mathrm{im}(t)$ and $k_\mathrm{RF}(t)$ are expected to provide similar estimates in the regime $\tau_\mathrm{mol} < t \ll \tau_\mathrm{rxn}$.
We note Eq.~\ref{equation:implied-rate} can be rearranged to yield a correlation function
\begin{eqnarray}
\frac{\expect{\delta h_A(0) \delta h_A(t)}}{\expect{\delta h_A^2}} &=& e^{-k_\mathrm{im}(t) \, t} 
\end{eqnarray}
By the definition of reactive flux (Eq.~\ref{equation:reactive-flux-correlation}), we can write $k_\mathrm{RF}(t)$ in terms of $k_\mathrm{im}(t)$ as,
\begin{eqnarray}
k_\mathrm{RF}(t) &=& - \frac{d}{dt} \frac{\expect{\delta h_A(0) \delta h_A(t)}}{\expect{\delta h_A^2}} = - \frac{d}{dt} e^{-k_\mathrm{im}(t) \, t} \nonumber \\
&=& e^{-k_\mathrm{im}(t) \, t} \left[ k_\mathrm{im}(t) + t \frac{d}{dt} k_\mathrm{im}(t) \right] 
\end{eqnarray}
When $t \gg \tau_\mathrm{mol}$, then $k_\mathrm{im}(t) \approx k$, and we have $k_\mathrm{RF}(t) \approx k e^{-k \, t}$.
 
\emph{Application of implied rate theory to single-molecule data.}
To illustrate the estimation of the phenomenological rate $k$ using the implied timescale $k_\mathrm{im}(t)$, we computed it for the p5ab hairpin force trajectory described above.
At the 50 kHz sampling rate (Fig.~\ref{figure:p5ab-hairpin}, right), the rate estimates are almost identical to those from $k_\mathrm{RF}(t)$ for a broad range of times where $t > \tau_\mathrm{mol}$, though there is much less noise in the $k_\mathrm{im}(t)$ rate estimate than in $k_\mathrm{RF}(t)$ (Fig.~\ref{figure:p5ab-hairpin}, right inset). 
At the 1 kHz sampling rate (Fig.~\ref{figure:p5ab-hairpin}, left), however, the rate estimate from $k_\mathrm{im}(t)$ remains stable over several times $\tau_\mathrm{rxn}$, even though the $k_\mathrm{RF}(t)$ has already decayed from the plateau region.
The implied rate estimate, $k_\mathrm{im}(t)$, therefore appears to provide a more robust estimate of the phenomenological rate under a variety of conditions. 

This robustness also carries over to an insensitivity to the placement of dividing surface $x^\ddag$, the problem reactive flux theory was originally envisioned to solve.
Using an observation time of $\tau = 60$ ms, the implied rate estimate $k_\mathrm{im}(\tau)$ varies much less than the na\"{i}ve rate estimates over a large range of dividing surface choices (Fig.~\ref{figure:p5ab-dividing-surface}, middle panels, black dashed and solid lines).

\emph{Microscopic rate constants.}
To obtain individual microscopic rates $k_{A \rightarrow B}$ and $k_{B \rightarrow A}$, we recall that the phenomenological rate $k$ represents the sum of the forward and backward rates,
\begin{eqnarray}
k &=& k_{A \rightarrow B} + k_{B \rightarrow A}
\end{eqnarray}
as well as the fact that the flux across the dividing surface must be balanced at equilibrium,
\begin{eqnarray}
\pi_A k_{A \rightarrow B} &=& \pi_B k_{B \rightarrow A}
\end{eqnarray}
which allows us to deduce that the individual rates are simply
\begin{eqnarray}
k_{A \rightarrow B} = \pi_B \, k \:\: ; \:\: k_{B \rightarrow A} = \pi_A \, k 
\end{eqnarray}
The equilibrium probabilities $\pi_A$ and $\pi_B$ can be simply estimated by the fraction of samples observed on each side of the dividing surface $x^\ddag$, such that $\pi_A \approx \expect{h_A}$ and $\pi_B \approx \expect{h_B}$.
For the RNA hairpin,  estimates of $\pi_A$ and $\pi_B$ are shown as a function of dividing surface placement in Fig.~\ref{figure:p5ab-dividing-surface} (bottom panel).
As both the equilibrium probability and phenomenological rate estimates are sensitive to the choice of dividing surface, the microscopic rates $k_{A \rightarrow B}$ and $k_{B \rightarrow A}$ will be \emph{more} sensitive to the dividing surface placement than either property alone.



The sensitivity of rates to the choice of dividing surface has some important implications.  
While thermodynamic quantities (e.g.~the free energy difference between two macrostates) are rather insensitive to the choice of dividing surface (as slight variation in $\pi_A$ and $\pi_B$ is suppressed by the logarithm in $\Delta G = - k_B T \ln (\pi_A / \pi_B)$), rates (and other kinetic properties such as commitment probabilities~\cite{chodera-pande:2011:prl:single-molecule-pfold}) typically have exponential weighting working in the opposite direction, making the definition of the surface particularly important.  
A key implication of this sensitivity is the challenge of comparing theory and experiment in kinetics---both must agree on the definition of the dividing surface in order to avoid confounding the comparison.
This is also of course an issue with even comparing different experiments.   
While this problem is unavoidable, our hope is that an approach which directly considers a detailed state decomposition~\cite{pande-beauchamp-bowman:methods:2010:msm-review,noe:jcp:2011:msm-review} will help further aid in the connection between theory and experiment.

\section*{Acknowledgments}

The authors would like to thank Ken Dill (University of California, San Francisco), Phillip L. Geissler (University of California, Berkeley), and Jed W.~Pitera (IBM Almaden Research Center) for stimulating discussions on this topic, and Gregory R.~Bowman (University of California, Berkeley) for helpful feedback on the manuscript.
PJE would like to thank Steve Smith (University of California, Berkeley) for help with the instrumentation and Jin Der Wen (National Taiwan University) and Ignacio Tinoco (University of California, Berkeley) for providing the p5ab RNA hairpin.  
This work was supported in part by NIH grants GM 32543 (C.B.), GM 50945 (S.M.) and a grant from the NSF (S.M.).
JDC gratefully acknowledges support from the HHMI and IBM predoctoral fellowship programs, NIH grant GM34993 through Ken A. Dill (UCSF), and NSF grant for Cyberinfrastructure (NSF CHE-0535616) through Vijay S. Pande (Stanford), and a QB3-Berkeley Distinguished Postdoctoral Fellowship at various points throughout this work.
FN acknowledges support from DFG Research Center Matheon.


\bibliographystyle{apsrev4-1} 
\bibliography{rate-theory}

\end{document}